\documentstyle{aipproc}
\begin{document}
\def\prb{Phys. Rev. B}
\def\prl{Phys. Rev. Lett.}
\def\pla{Phys. Lett. A}
\def\pr{Phys. Rev.}
\def\be{\begin{equation}}
\def\ee{\end{equation}}
\def\ba{\begin{eqnarray}}
\def\ea{\end{eqnarray}}
\def\Cal{\cal}
\def\LSCO{La$_{2-x}$Sr$_x$CuO$_4$}
\def\LCO{La$_2$CuO$_4$}
\def\LSNiO{La$_{2-x}$Sr$_x$NiO$_{4+\delta}$}
\def\YBCO{YBa$_2$Cu$_3$O$_{7-\delta}$}
\def\BKBO{BaKBiO}
\def\BSCCO{Bi$_2$Sr$_2$CaCu$_2$O$_{8+\delta}$}
\def\C60{A$_x$C$_{60}$}
\def\LNSCO{La$_{1.6-x}$Nd$_{0.4}$Sr$_x$CuO$_{4}$}
\def\optimalLCO{La$_{1.85}$Sr$_{.15}$CuO$_4$}
\def\VO{V$_2$O$_3$}
\def\TMTSF{(TMTSF)$_2$X}
\def\ET{BEDT...}
\def\hty{high temperature superconductivity}
\def\Hty{High temperature superconductivity}
\def\hts{high temperature superconductors}
\def\htr{high temperature superconductor}
\def\qp{quasiparticle}
\def\qps{quasiparticles}
\def\ua{\uparrow}
\def\da{\downarrow}
\parindent=4mm
\title{Local Electronic Structure and High Temperature Superconductivity.}
\author{V.~J.~Emery$^1$ and S.~A.~Kivelson$^2$}
\address{
$^1$Dept. of Physics
Brookhaven National Laboratory
Upton, NY  11973-5000}
\address{
$^2$Dept. of Physics
University of California at Los Angeles
Los Angeles, CA 90095}
\maketitle

\begin{abstract}

It is argued that a new mechanism and many-body theory of superconductivity 
are required for doped correlated insulators.
Here we review the essential features of and the experimental support for
such a theory, in which the physics is driven by the kinetic energy.

\end{abstract}

\section{Introduction}

{\Hty}\cite{BM} is obtained by adding charge carriers into a highly-correlated 
antiferromagnetic insulating state. Despite the fact that there is a large 
``Fermi surface'' containing all of the pre-existing holes and the doped 
holes,\cite{arpes} it is impossible to understand the behavior of the system 
and, in particular, the origin of {\hty} unless the nature of the 
doped-insulating state is incorporated into the theory. In particular, the 
Fermi liquid theory of the normal state and the BCS theory of the 
superconducting state, which are so successful for conventional metals, were 
not designed for doped insulators,  and they do not apply to the {\hts}. 
(Section II.) Consequently it is necessary to develop a new mechanism and 
many-body theory of {\hty}.

In our view, the physics of the insulator and the doped insulator, including 
antiferromagnetism and superconductivity, is driven by a lowering of the
zero-point kinetic energy.\cite{pwa} This is well known for the antiferromagnetic 
state but, in addition, the motion of a single hole in an antiferromagnet 
is frustrated because it stirs up the spins and creates strings of 
ferromagnetic bonds. Consequently, a finite density of holes forms self-organized 
structures, designed to lower the zero-point kinetic energy. This is accomplished in 
three stages: a) the formation of charge inhomogeneity (stripes), b) the 
creation of local spin pairs, and c) the establishment of a phase-coherent 
high-temperature superconducting state. The zero-point kinetic energy is 
lowered along a stripe in the first stage, and perpendicular to the stripe in 
the second and third stages.

Static or dynamical charge inhomogeneity,\cite{zaan,losala,ute,chayes,erica1} or 
``topological doping'' \cite{topo} is quite common for doped correlated insulators. 
In $d$ dimensions, the charge forms one-dimensional arrays of $(d-1)$-dimensional 
structures that are also antiphase domain walls for the background spins. In 
$d=1$ there is an array of charge solitons,\cite{review} whereas, in $d=2$, 
there are linear ``rivers of charge'' (stripes) threading through
the antiferromagnetic background.\cite{losala,ute,chayes} In $d=3$ there 
are arrays of charged planes\cite{chayes,erica1}, as observed in the manganates.%
\cite{mang} These self-organized structures, which may be fluctuating or
form ordered or glass phases, are a consequence of the 
tendency of the correlated antiferromagnet to expel the doped holes, and they 
lead to a lowering of the zero-point kinetic energy. The theoretical arguments 
that lead to this picture will be summarized in Sec. III.

It is clear that any new many-body 
theory must be based on the local electronic structure and there are
strong indications of a link to {\hty}. 
First of all, in LSCO and YBCO the value of T$_c$ is inversely
proportional to the spacing between stripes in underdoped and optimally doped
materials.\cite{yamada,bb} Secondly, $\mu$SR experiments 
\cite{weidinger,niedermeyer} have found evidence for a phase in which
superconductivity coexists with a cluster spin glass. In {\YBCO}, the spin 
freezing temperature goes to zero when the superconducting T$_c$  is more 
than 50K. It is difficult to see how these two phases could coexist unless
there is a glass of metallic stripes dividing the CuO$_2$ planes into
randomly-coupled antiferromagnetic regions.
A new mechanism and many-body theory of superconductivity, based on local charge 
inhomogeneity has been developed,\cite{badmetals,nature,ekz,erica2} 
and there is  substantial experimental support for the overall picture, 
as described in subsequent sections. 

\section{BCS Many-Body Theory}

There are several reasons why the Fermi liquid theory of the normal state and 
the BCS theory of the superconducting state do not apply to the {\hts}:

1) In BCS theory, the superfluid density $n_s$ is given by all electrons 
in the Fermi sea, whereas, in the {\hts}, $n_s$ is proportional to the
density of doped holes. 

2) The outstanding success of BCS theory stems from the existence of sharp
{\qps}. However, an analysis of the temperature dependence of the resistivity
shows that the {\qp} concept does not apply to many synthetic metals, 
including the {\hts}.\cite{badmetals,pphmf} This idea is supported by 
angular resolved photoemission spectroscopy (ARPES)
which shows no sign of a normal-state quasiparticle peak near the points 
$(0,\pm \pi)$ and $(\pm \pi,0)$ where {\hty} originates.\cite{norman,darpes} 

3) If there are no quasiparticles, there is no Fermi
surface in the usual sense of a discontinuity in the occupation number
$n_{{\vec k}}$ at $T=0$. This undermines the very foundation of 
the BCS mean-field theory, which is a Fermi surface instability.

4) In BCS theory, pairing and phase coherence take place at the same 
temperature T$_c$, and a good estimate of T$_c$ is given by $\Delta_0 /2$, 
where $\Delta_0$ is the energy gap measured at zero temperature. However, this
criterion does not give a good estimates of T$_c$ for the {\hts}, especially 
for underdoped materials: $\Delta_0 /2 T_c$ varies with doping and can be 
much greater than one. Rather, the value of T$_c$ is determined by the onset 
of phase coherence\cite{nature,erica2} and is governed by the 
zero-temperature value of the ``phase stiffness'', 
$V_0 \equiv (\hbar c)^2 a / 16\pi(e\lambda(0))^2$, which sets the energy 
scale for the spatial variation of the superconducting phase.
Here $\lambda(T)$ is the penetration depth and $a$ is a microscopic length scale 
that depends on the dimensionality of the material.\cite{nature}

5) A major problem for any mechanism of {\hty} is how to achieve a high pairing 
scale in the presence of the repulsive Coulomb interaction, especially in
a doped Mott insulator in which there is poor screening. In the {\hts}, the 
coherence length is no more than a few lattice spacings, so neither 
retardation nor a long-range attractive interaction is effective in 
overcoming the bare Coulomb repulsion. Nevertheless ARPES 
\cite{darpes} shows that the major component of the gap function is proportional to
$\cos k_x - \cos k_y$. It follows that, in real space, the gap function and
hence, in BCS theory, the net pairing force, is a maximum for holes 
separated by one lattice spacing, where the bare Coulomb interaction is very 
large ($\sim$ 0.5 eV, allowing for atomic polarization). It is not easy to 
find a source of an attraction that is strong enough to overcome the Coulomb
force at short distances and achieve a high transition temperature {\it in a natural 
way} by the usual Cooper pairing.

Clearly there is a need for a new mechanism and many-body theory to explain
{\hty}.

\section{Topological Doping}

It is well known that the motion of a single hole in an antiferromagnet is 
frustrated by the creation of strings of broken bonds.\cite{lev} 
This idea is supported  by ARPES, which found that the bandwidth of a single
hole is controlled by the exchange integral $J$, rather than the hopping 
amplitude $t$.\cite{wells} 

When there is a finite density of holes, the system strives to relieve this
frustration and lower its kinetic energy. If the holes were neutral the system 
would separate into a hole-free antiferromagnetic phase and a hole-rich 
(magnetically disordered or possibly ferromagnetic) phase, in which the holes 
are mobile and the cost in exchange energy is less than the gain in kinetic 
energy.\cite{ekl,marder,manousakis} In practice the holes are charged, but 
macroscopic phase separation can take place whenever the dopants are mobile, 
as in oxygen-doped and photo-doped materials. We have reviewed the experimental 
evidence for this behavior elsewhere.\cite{physicaC,losala} More recent 
experiments exploring oxygen doping in detail have been carried out by 
Wells {\it et al.}.\cite{wells2}

When the dopants are immobile, charged holes can do no more than phase separate locally, 
by forming arrays of linear metallic stripes \cite{losala,ute,chayes} which are 
``topological''  in nature, since they are antiphase domain walls for the 
antiferromagnetic background spins.\cite{topo,landau} This structure lowers 
the kinetic energy along the stripe but makes it more difficult, if anything, 
for a {\it single} hole to move perpendicular to the stripe direction. A hop 
transverse to a stripe takes the hole far above Fermi energy.\cite{ekz} 
However, as we shall see, {\it pairs of holes} can move more easily transverse 
to a stripe, and they lower their kinetic energy first by forming spin pairs
and, at a lower temperature, by making the system a {\htr}.

It has been argued that charge stripes are energetically impossible because
the driving energies are unable to overcome the Coulomb 
repulsion.\cite{phillips} However, charge modulation is {\it inevitable} if 
the short-range interactions give a negative compressibility, $\kappa$, as 
they do between the spinodals of a system that, otherwise, would undergo phase 
separation. A general expression for the Debye screening length is
$\lambda_D = \sqrt{\epsilon / 4 \pi e^2 n^2 \kappa}$, where $\epsilon$ is the dielectric
constant, $e$ is the charge and $n$ is the density. When $\kappa < 0$, $\lambda_D$
is imaginary, which indicates that the ground state is unstable to a density
modulation.\cite{pol,neto} Of course it requires a more detailed microscopic 
calculation to obtain the physical length scale. 

The existence of charge and spin stripes in the {\LSCO} family was established 
in an elegant series of experiments on {\LNSCO} by Tranquada and 
co-workers.\cite{jtran} In a Landau theory of the phase transition,%
\cite{landau} the spin order parameter ${\vec S}_{\vec q}$ and the charge 
order parameter $\rho_{-\vec Q}$ first couple in third order 
(${\vec S}_{\vec q}\cdot{\vec S}_{\vec q} \hskip 0.1 cm \rho_{-\vec Q}$), 
so the ordering vectors must satisfy ${\vec Q}=2 {\vec q}$ or, in other words, 
the wavelength of the spin modulation is twice that of the charge modulation.  
This relation is found to be satisfied experimentally,\cite{jtran} and it 
implies that the charge stripes also form antiphase domain walls in the 
magnetic order, which gives the precise meaning of the concept of topological 
doping.\cite{topo} The observation of essentially ordered 
stripes allowed a study of the evolution of the spin and charge order 
parameters, which not only provided input into the mechanism of stripe 
formation by showing that they are charge driven, but also established that 
{\it inelastic incommensurate magnetic} peaks observed previously 
\cite{cheong} in {\LSCO} were produced by fluctuating stripes. 
Recently, inelastic incommensurate magnetic peaks have been observed in 
underdoped {\YBCO} by neutron scattering experiments,\cite{mook} 
thereby establishing that {\YBCO} and the {\LSCO} family have a common 
spin structure. 

By now, the prediction of metallic stripes\cite{losala,ute} has been confirmed in 
all families of materials in which extensive neutron scattering experiments have 
been performed (LSCO and YBCO). There is growing evidence of similar behavior in 
{\BSCCO}: preliminary neutron scattering experiments show incommensurate 
magnetic peaks, and there is ARPES evidence\cite{shenstripe} of spectral weight 
transfer associated with stripes.
Also, a calculation of the effects of stripes in ARPES 
experiments \cite{markku} produced regions of degenerate states and a flat 
section of the ``Fermi surface'' near $(0,\pm \pi)$ and $(\pm \pi,0)$, as 
observed experimentally.\cite{shen,anl,norman}

\section {Spin Pairing}

The existence of a cluster spin-glass state for a substantial range of doping 
in the {\hts}\cite{weidinger,niedermeyer} implies that the stripe dynamics is
slow and that the motion of holes along the stripe is much faster 
than the fluctuation dynamics of the stripe itself. Thus an individual stripe 
may be regarded as a finite piece of one-dimensional electron gas (1DEG) 
located in an active environment of the undoped spin regions between the 
stripes. Then it is appropriate to start out with a discussion of an extended 
1DEG in which the singlet pair operator $P^{\dagger}$ may be written
\begin{equation}
P^{\dagger} = \psi^{\dagger}_{1 \ua} \psi^{\dagger}_{2 \da}  -
\psi^{\dagger}_{1 \da} \psi^{\dagger}_{2 \ua},  
\end{equation}
where $\psi^{\dagger}_{i, \sigma}$ creates a right-going ($i=1$) or
left-going ($i=2$) fermion with spin $\sigma$. In one dimension, the fermion 
operators of a 1DEG may be expressed in terms of Bose fields and their 
conjugate momenta ($\phi_c(x), \pi_c(x)$) and 
($\phi_s(x), \pi_s(x)$) corresponding to the charge and spin collective
modes respectively. In particular, the pair operator $P^{\dagger}$ becomes
\cite{review}
\begin{equation}
P^{\dagger} \sim e^{i\sqrt{2 \pi}\theta_c} \cos\big(\sqrt{2 \pi} \phi_s\big),
\end{equation}
where $\partial_x\theta_c \equiv \pi_c$. In other words, there is an 
{\it operator} relation in which the amplitude of the pairing operator depends 
on the spin fields only and the (superconducting) phase is a property of the 
charge degrees of freedom. Now, if the system acquires a spin gap, the 
amplitude $\cos\big(\sqrt{2 \pi} \phi_s\big)$ acquires a finite expectation
value, and superconductivity will appear when the charge degrees of freedom
become phase coherent. Clearly, in one dimension, the temperature at which the
spin gap forms is generically distinct from the phase ordering temperature 
because phase order is destroyed by quantum fluctuations, even at zero 
temperature.\cite{review}

We emphasize that we are {\it not} dealing with a {\it simple} 1DEG, for 
which a spin gap occurs only if there is an attractive interaction in the
the spin degrees of freedom.\cite{review} The 1DEG on the stripe is in
contact with an active (spin) environment, and we have shown that pair 
hopping between the 1DEG and the environment will generate 
a spin gap in both the stripe and the environment, even for purely 
repulsive interactions.\cite{ekz} The same mechanism gives rise to spin
gaps in spin ladders. Also, although the theory was worked out for an 
{\it infinite} 1DEG in an active environment, it is known from numerical
calculations on finite-size systems that the conclusions are correct for
any property that has a length scale small compared to the size of the system.
Here, we use the theory only to establish the existence of a spin gap, which
corresponds to a length scale of a few lattice spacings. Once a spin gap has 
been formed, the problem is reduced to the physics of the
superconducting phase and its quantum conjugate (the number density), and 
{\hty} emerges when phase order is established.\cite{badmetals,nature,erica2}

Experimentally the formation of an amplitude of the order parameter is 
indicated by a peak in $(T_1T)^{-1}$ (where $T_1$ is the spin-lattice 
relaxation rate),\cite{julien} 
and by ARPES,\cite{arpesgap} both of which are consistent 
with spin pairing. A drop in the specific heat\cite{loram} and a pseudogap in 
the $c$-axis optical conductivity,\cite{homes}
both of which indicate that the charge is involved, occur at a higher
temperature in underdoped materials, and are symptoms of the onset of stripe 
correlations.\cite{ekz}

\section{Phase Coherence}

{\Hty} is established when there is coherent motion of a pair from stripe
to stripe.\cite{ekz} This final step in the reduction of the zero-point 
kinetic energy is equivalent to establishing phase order, and it determines 
the value of T$_c$, especially in underdoped and optimally doped 
materials.\cite{badmetals,nature,erica2}

It is sometimes argued that thermal phase fluctuations are excluded because
the  Coulomb interaction moves them up to the plasma frequency, $\omega_p$, 
via the Anderson-Higgs mechanism. This argument, if correct, also 
would imply that critical phenomena near to T$_c$ cannot display 3d-XY
behavior. An explicit calculation shows why this objection is incorrect. The 
Fourier transform of the Lagrangian density in the long wavelength limit has 
the form\cite{badmetals,wagen}
\be
{\cal L}({\vec k}, \omega)=
{1 \over 2} {\vec k}^2 a^2 \big [V_0(\omega) + V_1 \omega^2 \epsilon(\omega)\big]
\theta^2({\vec k, \omega})
\ee
where $\epsilon(\omega)$ is the dielectric function at ${\vec k} = 0$,
$V_1 = \hbar^2 a/16 \pi e^2$, and
$V_0(0)\equiv V_0$ is the classical phase stiffness, defined above.
At high frequency 
\be
\epsilon(\omega) = \epsilon_{\infty}-{\omega_p^2 \over \omega^2}.
\ee
Note that $V_0(\omega)$ vanishes at high frequency and, in general, 
it does not contribute to the plasma frequency. Then phase fluctuations  
occur at a frequency $\omega_p/\sqrt{\epsilon_{\infty}}$. At the same time, 
for $\omega = 0$, the Lagrangian has the form ${\cal L} = const. {\vec k}^2$,
as required for classical phase fluctuations. In general, it is necessary
to do a renormalization group calculation to obtain the zero-frequency limit,
and we have shown that, for sufficiently good screening (large dielectric 
function), the behavior of the system is given by the classical ($V_0$) part 
of the Lagrangian.\cite{badmetals}  The unusual form of the Lagrangian stems
from the use of the dual phase-number representation, in which the $V_0$ 
term represents the kinetic energy (pair hopping) and the $V_1$ term is the
potential energy (Coulomb interaction).

We have solved the following model of classical phase 
fluctions\cite{nature,erica2}:
\be
H=  -J_{\parallel} \sum_{<ij>_{\parallel}} \left\{ 
\cos(\theta_{ij}) 
+ \delta\cos(2\theta_{ij})\right \} \nonumber  
- \sum_{<kl>_{\perp}} \left\{
J_{\perp}^{kl}\cos(\theta_{kl}) \right \},
\ee
where the first sum is over nearest neighbor sites within each plane, 
and the second sum is over nearest neighboring planes. The values of the 
constants $J_{\parallel}$ and $\delta$ are taken to be isotropic within 
each plane and the same for every plane. The coupling between planes, 
$J_{\perp}^{kl}$, is different for crystallographically distinct pairs of 
neighboring planes. 

The results of this final stage of the calculation are in good 
agreement with experiment. For a 
reasonable range of parameter values (as constrained by the magnitudes of the
penetration depths in different directions) the model gives a good estimate of 
T$_c$ and its evolution with doping. It also explains\cite{erica2} the temperature 
dependence of the superfluid density, obtained for a range of doping by 
microwave measurements.\cite{hardy} 

The phase diagram itself is consistent with this picture. The physics evolves 
in three stages. Above the superconducting transition temperature there are 
two crossovers, which are quite well separated in at least some underdoped 
materials. The upper crossover is indicated by the onset of short-range 
magnetic correlations and by the appearance of a pseudogap\cite{homes} in the 
$c$-axis optical conductivity (perpendicular to the CuO$_2$ planes) which 
might possibly indicate the establishment of a stripe glass phase. The lower 
crossover is where a spin gap or pseudogap (which is essentially the 
amplitude of the superconducting order parameter) is formed.
Finally, superconducting phase order is established at T$_c$ and, in fact,
determines the value of T$_c$.\cite{nature,erica2}

{\bf Acknowledgements:}  We acknowledge frequent discussions 
with J.~M.~Tranquada.
This work was supported at UCLA by the National Science Foundation grant 
number DMR93-12606 and, at Brookhaven,  by the Division of Materials Sciences,
U. S. Department of Energy under contract No. DE-AC02-98CH10886.

\end{document}